# Clifford´s Attempt to test his Gravitation Hypothesis

S. Galindo and Jorge L. Cervantes-Cota
Departamento de Física, Instituto Nacional de Investigaciones Nucleares
Km. 36.5 Carretera México-Toluca 52045, México
e-mails: salvador.galindo@inin.gob.mx
jorge.cervantes@inin.gob.mx

Casi medio siglo antes de que Einstein expusiera su teoría general de la relatividad, el matemático inglés William Kingdon Clifford argumentó que el espacio podría no ser euclidiano y propuso que la materia no es más que una pequeña distorsión en esa curvatura espacial. Propuso además que la materia en movimiento no es más que la simple variación en el espacio de esta distorsión. En este trabajo, conjeturamos que Clifford fue más allá de dichas propuestas, pues trató de demostrar que efectivamente la materia curva el espacio. Para ello hizo una observación infructuosa sobre el cambio del plano de polarización de la luz solar durante el eclipse solar del 22 de diciembre de 1870 en Sicilia.

Descriptores: W K Clifford; Prueba de Teorías Gravitacionales; Historia de la Gravitación; Historia de la Relatividad General

Almost half a century before Einstein expounded his general theory of relativity, the English mathematician William Kingdon Clifford argued that space might not be Euclidean and proposed that matter is nothing but a small distortion in that spatial curvature. He further proposed that matter in motion is not more than the simple variation in space of this distortion. In this work, we conjecture that Clifford went further than his aforementioned proposals, as he tried to show that matter effectively curves space. For this purpose he made an unsuccessful observation on the change of the plane of polarization of the skylight during the solar eclipse of December 22, 1870 in Sicily.

**Clifford´s Attempt to Test his Gravitation Hypothesis**

1. Introduction

During the course of the eighteenth century, brilliant mathematicians had shown that Newton´s law of Gravitation accounted for many astronomical and terrestrial phenomena. They proved that Universal Gravitation was capable of accurately describing the observed motions of celestial bodies. As result, Universal Gravitation came to be at that time, the archetype of scientific theories, the shining example of Physics.

However, from its initial conception, Newton explicitly refused to discuss the cause of gravitation. "*I frame no hypotheses*" (*Hypothesis non fingo)*, he wrote in the *General Scolium* which he appended to the second edition of his *Principia* [1]. Moreover, he considered that the cause of mutual attraction of matter was not assigned to the epistemological scrutiny of physics. In fact, in the same *General Scolium* he openly asserted,

*"I have not been able to discover the cause of those properties of gravity from phenomena, and I frame no hypotheses; for whatever is not deduced from the phenomena is to be called an hypotheses; and hypotheses, whether metaphysical or physical, whether of occult qualities or mechanical, have no place in experimental philosophy. In this philosophy particular propositions are inferred from the phenomena, and afterwards rendered general by induction."* [1]

Hence, in the years to come the essence of gravitation grew out to be unfathomable to many, as such issue was according to Newton no longer considered to be part of the proper investigation of Physics (i.e. *magister dixit*).

By the late eighteen century the notion of instant action at distance of ponderable matter acting upon other matter had become accepted as an irreducible concept by the overall majority of natural philosophers. In 1871, Ernst Mach (1838-1916) could still state that "*Nowadays gravitation does no longer perturb any human being. It has become an ordinary incomprehensibility*" (*Eine gewöhnliche unfassbarkeit*) [2]

However some scholars did not exclude a proposal on the origin of gravitation setting aside Newton´s "*hypothesis non fingo*". One of them was William Kingdon Clifford (1845-1879) who maintained the radical stance that gravitation may be explained in terms of the curvature of space and its changes [3].

In what follows we shall present and discuss Clifford´s proposal on the nature of gravitation and in our opinion what constitutes his attempt to prove the truth of his postulate.

## 2. An overview of Clifford´s vitae

William Kingdon Clifford (1845 - 1879) was in his days an outstanding mathematician, appreciated by his colleagues for his lectures and publications. Clifford was born in Starcross near Exeter England, where he received his first education. When he was fifteen, Clifford won a scholarship to London's King's College. Before leaving King´s at the age of eighteen, he had already published his first mathematical paper. In 1863 he went up to Trinity College, Cambridge. There he ranked second "*wrangler*" in Cambridge's famous mathematical examination, the "*tripos*". In 1867 he was awarded a B.A. degree in Mathematics and Natural Philosophy. Afterwards he was elected Fellow of Trinity at the early age of 23, and 3 years later despite of his youth, appointed Professor of Applied Mathematics and Mechanics at University College, London. He remained in this chair until his premature death from tuberculosis at the age of 33 [4].

After passing away, Clifford´s fame diminished but his work was not completely forgotten, he had some well-known adherents that carried on his ideas: Charles Sanders Peirce (1839 -1914) and Felix Klein (1849 -1925), to name two. However, it was not until the 1970s that his mathematical work was reappraised. A reason for this revival was his valuable development of a mathematical scheme that today is widely used in the theory of Particle Physics. This powerful mathematical apparatus is now known by the name of "Clifford algebra". But what concerns us here is Clifford´s interest in Riemann´s ideas on the nature of physical forces.

## 3. Riemann´s ideas in short.

Bernhard Riemann (1826-1866) is considered one of the most brilliant mathematicians of all times. However his interest in physical phenomena is less known. Felix Klein once said of him in 1894, that Riemann´s work was characterized by his continual attempt to put "*in mathematical form a unified formulation of the laws which lie at the basis of all natural phenomena*" [5]

Klein´s words are indeed accurate. Riemann´s himself manifested his interest in physics as early as November 1850. In that year he wrote an article announcing his confidence in the possibility of creating a coherent mathematical theory "*on a continuous medium*", which encompassed the phenomena of electricity, magnetism, light, heat balance and gravity [6]. In the same November 1850 article, Riemann maintained that it was possible to formulate a mathematical theory by moving from elementary principles toward general laws valid in all of a given continuous space. Moreover, in an unfinished manuscript dated March 1$^{st}$ 1853 just a year before delivering his habilitation lecture, he proposed the very audacious idea that the universe is filled with some kind of ether which he called "*Stoff*", that ran into atoms to fade away,

"*I make the hypothesis that space is filled with a substance which continually flows into the ponderable atoms where it disappears from the world of phenomena* "[7].

It is appropriate to say that in later writings Riemann preferred to use the term "ether" instead of "*Stoff*". Then in a different manuscript dated in the same 1853, he suggested as the cause of gravitation, the behavior of his hypothetical "*Stoff*" in an empty geometrical space.

*"I seek the cause [of gravitation] in the state of motion of the […] continuous substance (Stoff) spread throughout the entire infinite space. […] Thus this substance may be thought of as a physical space whose points move in the geometrical space."* [8]

Thus, in order to build a theory, Riemann had the idea of using two separate entities, namely: the continuous substance "Stoff" and the geometrical space. His idea on the geometrical space was to be revealed soon.

By 1853 Carl Friedrich Gauss (1777-1855) asked Riemann, at that time his student, to prepare a *Habilitationsschrift* on the foundations of geometry. Over many months, Riemann developed his theory of higher dimensions and delivered, on June 10th 1854 to the Philosophical Faculty of the University of Göttingen, his Habilitation lecture entitled "*Ueber die Hypothesen welche der Geometrie zu Grunde liegen*" [9]. Initially in this dissertation, Riemann mentioned that a space is a continuum in which points are indicated by the values of their *n* coordinates, *n* being the dimension of the space. Then he argued that such space might have one of several possible geometric structures. Just before the end of his dissertation, Riemann developed the idea that space, even though of Euclidean appearance at macroscopic scales, may well have a non-Euclidean geometric structure from the viewpoint of variable curvature.

Riemann´s *Habilitationsschrift* was only published twelve years later in 1868 (two years after Riemann´s death) by Richard Dedekind (1831-1916). Today this work is recognized as one of the most fundamental works in geometry. However many years before, Immanuel Kant (1724-1804) maintained that the Euclidean character of space was a fact of nature, and the high-ranking weight of his opinion was for long time dominant among contemporary scholars. As consequence, the early reception of Riemann´s paper was slow because the widespread view that there was no other geometry but Euclidean. In Hermann Weyl´s (1885-1955) words, Riemann´s Habilitation essay *"was not grasped by his contemporaries, and his words died away almost unheard (with the exception of a solidary echo in the writings of W.K. Clifford)"* [10].

William Clifford was Riemann´s follower. In fact, he made the first translation into English of Riemann's 1854 Habilitation paper on the new non-Euclidean geometries under the title "*On the hypotheses which underlie geometry*" [11]. Furthermore, following Riemann's ideas, Clifford developed, as we shall next see, a proposal which extended the belief that the intrinsic nature of space is non-Euclidean and that gravitation could be formally represented by this underlying geometry.

### 4. Clifford´s Gravity proposal.

On the 18th of August 1869, the eminent mathematician James Joseph Sylvester delivered his opening presidential address to the Mathematical and Physical section of the British Association at their meeting in Exeter [12]. His talk, entitled "*A Plea for the Mathematician*", was an ingenious response to Prof. Thomas Henry Huxley previous remarks made in the course of an after-dinner speech delivered before the Liverpool Philomathic [sic] Society in April 1869. This speech was afterwards published in MacMillan´s Magazine [13].

In his Liverpool speech, Huxley emphasized only one particular aspect of the nature of Mathematics, namely: "*mathematical training is purely deductive*" and went on asserting "*The mathematician starts with a few simple propositions, the proof of which is so obvious that they are called self-evident, and the rest of his work consists of subtle deductions from them*"[13].

Later Huxley´s position became more radical in an essay written for the June 1869 issue of the Fortnightly Review. There he wrote that Mathematics "*is that which knows nothing of observation, nothing of experiment, nothing of induction, nothing of causation*" [14].

The remarks made by Huxley (who was at the time president-to-be of the British Association) sparkled Sylvester´s quick temper and as consequence he prepared a rebuttal delivered at the above mentioned Exeter meeting [12].

In his response Sylvester made emphasis on some other facets of Mathematics ignored by Huxley, those were: the nature of mathematical induction and on practical and applied aspects of the discipline. Sylvester brought to mind that "*…One of the principal weapons* [of Mathematics] *is induction, that it has frequent recourse to experimental trial and verification and that it affords a boundless scope for the exercise of the highest efforts of imagination and invention*"[12].

In allusion to Huxley´s aforementioned reproach "… [Mathematics] *knows nothing of observation*" Sylvester wrote,

"*Gauss has called Mathematics the science of the eye... Riemann has written a thesis to show that the basis of our conception of space is purely empirical,…that other kinds of space might be considered to exist…Like his master Gauss, Riemann refuses to accept* **Kant´s doctrine of space and time being forms of intuition**, *and regards them as possessed of physical and objective reality*". [12], (emphasis is ours).

Sylvester then drew attention to the possibility of a raised dimensional space and for this purpose he used an analogy to explicate this speculated experience: "*For as we can conceive beings (like infinitely attenuated bookworms in an infinitely thin sheet of paper) which possess only the notion of space of two dimensions, so we may imagine beings capable of realizing space of four or a greater number of dimensions*" [12].

Sylvester´s presidential address was afterwards published and appended with supplemental footnotes in *Nature* that at the time was a new periodical. One of the footnotes in Sylvester´s address disclosed Clifford´s ongoing groundbreaking work. The footnote reads,

"[ … ] *the laws of motion accepted as fact, suffice to prove in a general way that the space we live in is a flat or level space (a "homaloid")* [i.e. Euclidean], *our existence therein being assimilable* [i.e. similar] *to the life of the bookworm in a flat page; but what if the page should be undergoing a process of gradual bending into a curved form?*" [12].

Then Sylvester made the first public announcement that Clifford had been working on a new concept of space,

*"Mr. W.K. Clifford has indulged in more remarkable speculations as the possibility of our being able to infer, **from certain unexplained phenomena of light** and magnetism, the fact of our level space of three dimensions being in the act of undergoing in space of four dimensions (space as inconceivable to us as our space to the supposititious bookworm) a distortion analogous to the rumpling of the page."*[12], (emphasis is ours).

Sylvester´s publication triggered an epistolary discussion among Nature´s readership for the most part on Sylvester´s interpretation of Kant´s doctrine of space and time. The reaction produced a series of letters to *Nature*´s Editor, published in the pages of the same Journal. The letters approved or objected Sylvester´s opinion on Kant´s vison of space. In a short time, the epistolary skirmish escalated drawing in correspondents both within and outside of the academia. By January 29 of 1870, Norman Lockyer, editor of *Nature* decided to end the epistolary exchange in his journal by publishing an editorial letter stating: "*this correspondence must now cease*". Here we will not deal with this matter any further. However there was a pair of letters, dated January 13 and February 17 1870, by a Mr. C. M. Ingleby expressing disagreement with Clifford´s conviction that the intrinsic nature of space is non-Euclidean [15, 16].

Clifford on the other hand, decided not to reply to Ingleby´s letters to Nature but instead, few days later (February 21st 1870) he publically exposed his ideas in a lecture entitled "On the Space-Theory of Matter" read to the Cambridge Philosophical Society. There Clifford maintained his radical stance that gravitation may be explained in terms of the curvature of space and its changes. Unfortunately his seminal lecture was never completely published, but in spite of this, we can get a good idea of its content as a resume is available nowadays. In it the 24-year old wrote, [17].

"*I wish here to indicate a manner in which these speculations* [namely Reimann´s i.e., that there are different kinds of space and that we can find out by experience to which of these kinds of space in which we live belongs] *may be applied to the investigation of physical phenomena.*

*I hold in fact:*

*1. That small portions of space are in fact of a nature analogous to little hills on a surface which is on the average at; namely, that the ordinary laws of geometry are not valid in them.*

*2. That this property of being curved or distorted is continually being passed from one portion of space to another after the manner of a wave.*

*3. That this variation of the curvature of space is what really happens in that phenomenon which we call the motion of matter, whether ponderable or etherial.*

*4. That in the physical world nothing else takes place but this variation, subject (possibly) to the law of continuity."*

It is interesting that in the second point of this list, Clifford holds the existence of some sort of wave that continually conveys space curvature from one spatial point to another, yet this brief mention of a wave does not indicate that he had any thoughts about propagating ripples as a class of self-propagating waves.

Immediately after presenting the list of Clifford´s four statements, the brief resume adds an apparently out of context paragraph but crucial for our discussion that follows,

*"I am endeavouring in a general way to explain the **laws of double refraction** on this hypothesis, but have not yet arrived at any results sufficiently decisive to be communicated".* [17], (emphasis is ours).

If we bring to mind what Sylvester had just recently expressed (August 1869) in his Exeter speech (that Clifford was working on "certain unexplained phenomena of light,"), the preceding paragraph indicates that Clifford was trying to explain ( by February 1870) a phenomena linked to double refraction on grounds of his hypothesis of curved space. This raises the specific question on, what was this unexplained phenomenon supposedly related to double refraction? In the next section we will show that at the time, skylight polarization was believed to be associated with double refraction. Therefore, in our view, Clifford´s interest was on the space curvature influence on this phenomenon. But then, in his own words, he had not yet arrived to "any results sufficiently decisive". In our opinion, this prompted him to seek a direct clue that proved the link of skylight polarization with the curvature of space. Clifford may have asked himself; how might I find this clue? Before long, his chance was produced.

In that very month of February 1870 (Clifford´s Cambridge Society lecture was given the 21$^{st}$ of February), a communication in *Nature*'s "*notes section*" appeared, announcing that the Royal Society and the Royal Astronomical Society have appointed "*committees of council*" to report upon the steps to take in connection to the next total solar eclipse, visible in Algeria, Spain and Sicily during December of that very year [18]. Not surprisingly Clifford was going to be one of the members of the English expedition party to observe that eclipse. In particular Clifford´s task was to measure the effect of the eclipse in skylight polarization, before, during and after the event [19]. This, we must accept, was an odd activity for a Mathematician. In fact Clifford´s undertaking during the expedition was analogous to Eddington´s 1919 famous eclipse observation, but in the latter, the aim was to measure the bending of light by our Sun, while in the former, a possible change in the polarization plane by the moon´s supposedly distortion of space.

Before considering the outcome of the English Solar Eclipse Expedition of 1870, we shall now turn our attention to learn how the term "double refraction" was associated to skylight polarization in mid-XIX century.

5. **The polarization of the Skylight**

In December 1868, just few months before Sylvester´s address to the Cambridge society (delivered in August 1869), the prominent physicist John Tyndall (1820-1893) admitted,

*"The blue colour of the sky, and the polarization of skylight …constitute, in the opinion of the most eminent authorities, **the two great standing enigmas** of meteorology "*, [20] (emphasis is ours).

In effect, by mid nineteen century skylight polarization was a mystery among the scientific community even though many characteristics of the phenomenon had already been gathered and many explanations were brought forward, but none of them were satisfactory.

In 1809 Francois Arago discovered that light in the sky itself was partially polarized in a direction tangential to a circle centered in the sun and he found that maximum polarization is located at right angles to the Sun [21]. When he tried to identify a polarization pattern he observed a spot in the sky where the skylight polarization switches from vertical to horizontal going from the horizon to the zenith. This point is nowadays called the Arago neutral point (see figure 1). In 1840 the French meteorologist Jacques Babinet located a second neutral point situated above the Sun [22] Since a neutral point existed above the Sun, from considerations of symmetry, the Scottish physicist David Brewster predicted, years later, a third point of zero polarization positioned at a similar angular distance below the Sun along the solar meridian. This celestial point, known as the Brewster neutral point, was found in 1842 at the theoretically predicted position [23]. Only in 1846 could Babinet confirm the existence of the Brewster point [24].

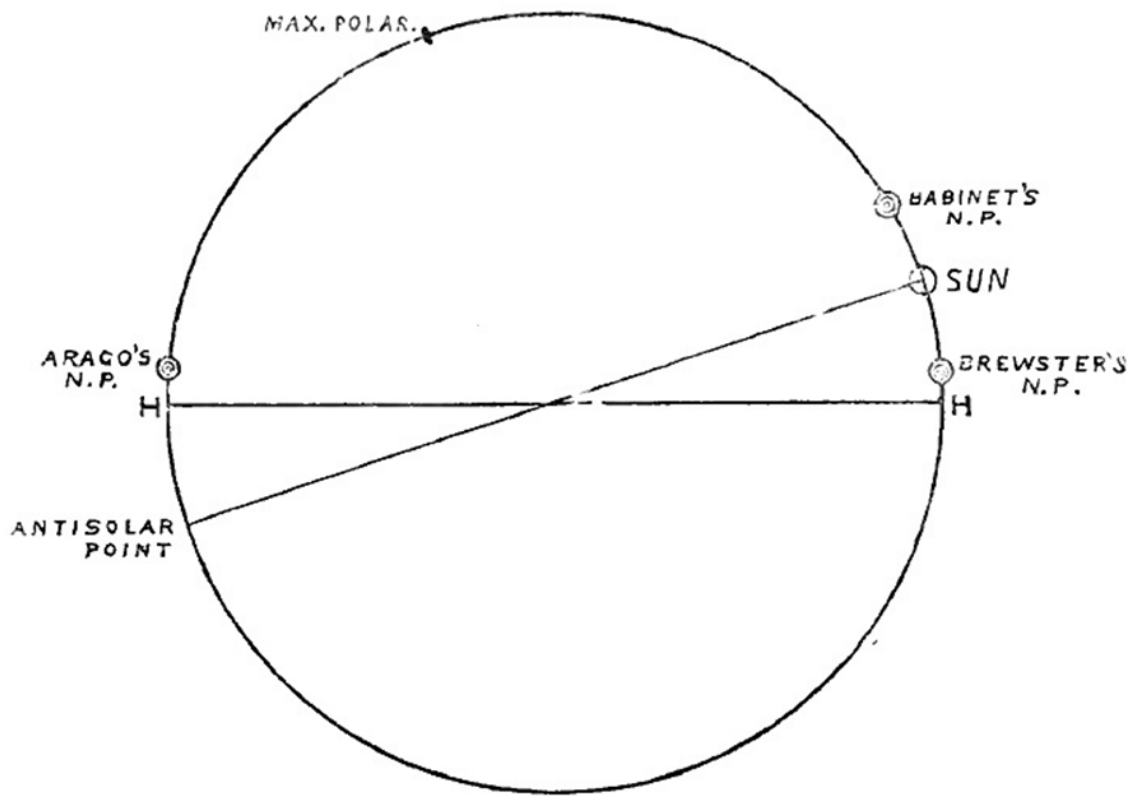

Figure 1. *Celestial sphere and horizon line (H –H) showing three positions where the skylight is unpolarized: the Arago, Babinet, and Brewster neutral points (N.P.). Taken from Brewster´s paper* [24]

By the mid nineteen century sky surveillances have already confirmed three sites in the celestial sphere where the skylight is unpolarized, namely the already mentioned: Arago, Babinet, and Brewster neutral points (Fig. 1). This polarization pattern was baffling since the simplest single scattering theory predicts a sky without the mentioned neutral points and the region opposite to the Sun, completely unpolarized. Nowadays these unpolarized points are interpreted as a multiple Rayleigh scattering effect. However, physicists of that time were puzzled. Instead of finding an explanation on what could possibly explicate the polarization of the sky, they were discarding hypotheses.

For example, Herschel ruled out single refraction as the cause of the observed pattern. In a letter written in 1862 to Tyndall we can read.

"*The cause of polarization is evidently a reflection of the sun´s light upon "something", the question is, on what? Were the angle of maximum polarization is 76$^o$, we should look to water or ice as the reflecting body, however inconceivable the existence in a cloudless atmosphere in a hot summer´s day of unevaporated molecules (particles?) of water. But though we were once of this option, careful observation has satisfied us that 90$^o$* (see Fig.1), *or thereabouts, is the correct angle, and that therefore, whatever be the body on which the light has been reflected, if polarized by a single reflection, the polarizing angle must be 45$^o$, and the index of refraction, which is the tangent of that angle, unity; in other words, the reflection would require to be made in air upon air*". [25]

Meanwhile Brewster surveyed polarization by reflection of light by rough and white surfaces. He worked "*under the conviction that the sky or atmosphere was a rough surface like any aggregation of white or coloured particles*" [26]. His results were not convincing and he concluded in 1863 that,

"*It is not one of the least wonders of terrestrial physics, that the blue atmosphere which overhangs us exhibits in the light which it polarizes phenomena somewhat analogous to those of crystals with two axes of **double refraction***", [ref 26 p. 210] (emphasis is ours).

In this way Brewster´s manuscript coined the term "*double refraction*" for what today is known as the skylight polarization. This is the reason why, when Clifford writes in his multi-quoted paragraph "*to explain the laws of double refraction on this hypothesis",* he actually means to explain the laws of skylight polarization on the hypothesis of the curvature of space.

### 6. The December 22$^{nd}$ 1870 total solar eclipse expedition.

The English Eclipse Expedition set off earlier in December 1870, on the steamship H.M.S. Psyche scheduled for a stopover at Naples before continuing to Syracuse in Sicily. Unfortunately before arriving to her final call, the ship struck rocks and was wrecked off Catania. Fortunately all instruments and members of the party were saved without injury [19].

Originally it was the intention of the expedition to establish in Syracuse their head-quarters, but in view of the wreckage the group set up their base camp at Catania. There the expedition split up into three groups. The group that included Clifford put up an observatory in Augusta near Catania.

The leader of this group was William Grylls Adams professor of Natural Philosophy at King's College, London [19].

In a report written by Prof. Adams, describing the expedition, we learn that the day of the eclipse, just before the time of totality, "… *a dense cloud came over the Moon and shut out the whole, so that it was doubtful whether the Moon or the clouds first eclipsed the Sun … Mr. Clifford observed light polarized on the cloud to the right and left and over the Moon, in a horizontal plane through the Moon´s centre …It will be seen from Mr. Clifford´s observations that the plane of polarization by the cloud…was nearly at right angles to the motion of the Sun*" [19].

As was to be expected, Clifford´s eclipse observations on polarization did not produce any result His prime intention, of detecting angular changes of the polarization plane due to the curving of space by the Moon in its transit across the Sun´s disk, was not fulfilled. At most he confirmed the already known information, i.e. the skylight polarization plane moves at right angles to the Sun anti-sun direction.

**Aftermath**

Soon after his "unsuccessful" involvement in the solar eclipse expedition, Clifford returned to England to take a post as Professor of Applied Mathematics and Mechanics at University College, London in the spring of 1871. There he continued championing the possibility of space being curved as well as delivering lectures among popular audiences on the topic. These lectures were given as part of the afternoon lecture series at the Royal Institution on March 1st, 8th, and 15th of 1873. The lectures were subsequently printed under the titles: "The Philosophy of the Pure Sciences" [27], "The Postulates of the Science of Space" [28] and "The Universal Statements of arithmetic" [29].

During the course of these lectures Clifford explained to the public some of the consequences of inhabiting in a curved universe. For example, he described the effects of living in a Universe where its curvature is nearly uniform and positive. He pointed out that if that were the case then "*the Universe…for the extent of space, is a finite number of cubic miles*" [ref. 28, p.386]. In simple words, the universe might be boundless but finite. He also pointed out the outcome, in that Universe, of traveling an enormously long distance about a straight line,

"*If you were to start in any direction in a perfect straight line according to the definition of Leibnitz; after traveling a most prodigious distance, to which the parallactic unit – 2000,000 times the diameter of the earth´s orbit – would be only a few steps, you would arrive at--- this place* [the lecture room]- *Only if you had started upwards, you would appear below*" [ref. 28, p.387].

In addition, to his lecture notes, he partially wrote a book entitled "Common Sense in Exact Sciences" that he couldn´t finish as his health started to decline. However the latter book was completed and edited in 1885 by Karl Pearson six years after Clifford's unfortunate death [30].

After his early death at the age of 33, his papers and notes were gathered together by his friends and followers and some were published. During a search in Clifford´s note-books, a slip of paper was found with the following words in Clifford´s handwriting, "Force is not a fact al all, but an idea embodying what is approximately the fact". [ref. 30 p. ix footnote 2]

In effect, Clifford believed that dynamical forces in three-dimensional space reduced to kinematical motions in four-dimensional space. With the intention to formalize these ideas into a mathematical theory, Clifford developed his own mathematical system as Tensor Theory was still in its infancy. He produced his own formalism of four-dimensional biquaternions to describe Rimannian geometry so to reduce three-dimensional dynamics to the four-dimensional kinematics in an elliptic space. In 1878 he gathered together these ideas in his work "Elements of Dynamic books 1" [31]. Additional volumes (2, 3 and 4) were posthumously published [32].

**Conclusion**

In this work, we put forward the conjecture that Clifford tried to show that matter effectively curves space. For this purpose he made an unsuccessful observation on the change of the plane of polarization of the skylight during the solar eclipse of December 22, 1870 in Sicily. As we have already stated, Clifford´s undertaking was in a certain sense visionary since it is an analogous experience to that of Eddington´s 1919 famous eclipse observation. Except that in Clifford´s observation, the idea was detecting the bending of the polarization plane by the moon´s distortion of space, while in Eddington´s, the bending of light by our sun.

Nowadays Clifford is better known for his mathematical works: Clifford numbers, Clifford algebras, Clifford-Klein surfaces but is less known in physics, at most he is known by very few for having anticipated Einstein's curved space paradigm of general relativity. But he did more than that. Here, in this work, we have proposed the idea that he also tried to test this paradigm.